\documentclass[aps,prd,twocolumn,a4paper,floatfix,nofootbib,superscriptaddress,nofootinbib]{revtex4}
\usepackage{amsmath,amsfonts,amssymb}
\usepackage{graphicx}
\usepackage{calligra}
\usepackage{units}
\usepackage{natbib}
\usepackage{tikz,pgfplots}
\usetikzlibrary{decorations.markings,arrows.meta}

\DeclareMathAlphabet{\mathcalligra}{T1}{calligra}{m}{n}
\newcommand{\rth}{\mathrm{th}}
\newcommand{\re}{\mathrm{e}}
\newcommand{\rn}{\mathrm{n}}
\newcommand{\req}{\mathrm{eq}}
\newcommand{\hden}{\mathcalligra{h}\,\,}

\begin{document}

\title{Simulating bulk viscosity in neutron stars. I. Formalism}
\author{Giovanni \surname{Camelio}}
\affiliation{Nicolaus Copernicus Astronomical Center, Polish Academy of Science, Bartycka 18, 00-716 Warsaw, Poland}
\author{Lorenzo \surname{Gavassino}}
\affiliation{Nicolaus Copernicus Astronomical Center, Polish Academy of Science, Bartycka 18, 00-716 Warsaw, Poland}
\affiliation{Department of Mathematics, Vanderbilt University, Nashville, TN, USA}
\author{Marco \surname{Antonelli}}
\affiliation{CNRS/IN2P3, ENSICAEN, Laboratoire de Physique Corpusculaire, 14000 Caen, France}
\affiliation{Nicolaus Copernicus Astronomical Center, Polish Academy of Science, Bartycka 18, 00-716 Warsaw, Poland}
\author{Sebastiano \surname{Bernuzzi}}
\affiliation{Theoretisch-Physikalisches Institut, Friedrich-Schiller-Universit{\"a}t Jena, 07743, Jena, Germany}
\author{Brynmor \surname{Haskell}}
\affiliation{Nicolaus Copernicus Astronomical Center, Polish Academy of Science, Bartycka 18, 00-716 Warsaw, Poland}
\date{\today}

\begin{abstract}
The faithful inclusion of the effects of bulk viscosity induced by the presence
of chemical reactions is an important issue for simulations of core-collapse
supernovae, binary neutron star mergers, and neutron star oscillations, where
particle abundances are locally pushed out of chemical equilibrium by
rarefaction and compression of the fluid elements.
In this work, we discuss three different approaches that can be used to
implement bulk viscosity in general relativistic hydrodynamic simulations of
neutron stars: the exact multi-component reacting fluid, and two
M\"uller-Israel-Stewart theories, namely the second order Hiscock-Lindblom
model and its linear limit, the Maxwell-Cattaneo model.
After discussing the theory behind the three approaches, we specialize their
dynamics equations to spherical symmetry in the radial gauge-polar slicing
(i.e., Schwarzschild) coordinates.  We also discuss a particular choice for the
equation of state of the fluid and the associated neutrino emission rates,
which are used in a companion paper for the numerical comparison of the three
frameworks, and we obtain the effective sound speed for the Hiscock-Lindblom
theory in the non-linear regime.
\end{abstract}

\maketitle

\section{Introduction}
\label{sec:intro}

With the detection of gravitational and electromagnetic emission from a binary
neutron star merger \citep{AbbottNS1,AbbottNS2}, we are now in an era when
information on hot and dense nuclear matter can be extracted from the
observation of extreme astrophysical phenomena \citep{AbbottNS3}. With more
observations to become available in the near future, reliable numerical
relativity simulations will be necessary to interpret the data. For this, it is
fundamental that the theoretical models used in simulations can faithfully
reproduce the physics of the system. 

Until recently \citep[e.g.,][]{Duez04, Radice17, Shibata17, SK17, SK172,
Kiuchi18, RadiceEjecta, Perego19, Vigano20, Radice20, Shibata21}, most
simulations of neutron star mergers assumed that matter behaves as a perfect
fluid. However, several dissipative processes can take place in the merger remnant,
which is a hot, rotating, and massive neutron star.  While on one hand there is
a consensus on the importance of magneto-hydrodynamic viscosity in this phase
\citep{Kiuchi18, Radice17, Shibata17, RadiceEjecta, Vigano20, Radice20,
Shibata21}, on the other hand the role of bulk viscosity is more controversial:
order of magnitude arguments show that weak interactions\footnote{However,
reactions are not the only microscopic process that can give rise to bulk
viscosity \cite{UdeyRadiation1982,Hiscock91,GavassinoRadiazio2020}. For example, even simple (i.e., a single species with no internal
degrees of freedom) weakly relativistic ideal gases can display bulk viscosity if driven out of
thermodynamic equilibrium by a volume change \citep{landau10,
Gavassino21bulk}.} (such as direct \citep{haensel_bulkI} or modified
\citep{haensel_bulkII} Urca processes, that tend to re-establish the chemical
equilibrium after a perturbation \citep[e.g.,][]{Haensel_Bulk_Urca,
haensel_frozen_2002, Gusakov_kantor_2008, Schmitt_review_2018}) can in
principle affect the stellar dynamics \citep{Yakovlev_review_2001, Schmitt_review_2018,
AlfordRezzolla, Most22}, but no clear evidence of this effect has
been found in simulations \citep{Perego19, Most21, Hammond2021, Radice22}.

Viscosity is also known to be one of the dissipative processes that can
efficiently damp oscillations around equilibrium configurations of both hot and
cold neutron stars \citep{cutler1987, Sawyer_Bulk1989, andersson_jones_2000}. For example, r-modes
are unstable to gravitational wave emission, but this mechanism is suppressed
by the shear and bulk viscosity at low and high temperatures, respectively
(see, e.g., \citep{haskell_rmode_review_2015} for a review).

Bulk viscosity is present also in core-collapse supernovae, where huge amounts
of neutrinos (on the order of a tenth of a solar mass \cite{Burrows88}) are
released during the collapse.  As the chemical composition of matter and
neutrinos is of the utmost importance for the supernova explosion,
core-collapse supernova simulations routinely employ a multi-component fluid
description, see e.g.~\citet{OConnor10}.

In this paper, we discuss three different approaches to bulk viscosity (the
multi-component fluid \citep{carter1991, Gavassino21bulk}, Hiscock-Lindblom
\citep{Hiscock1983}, and Maxwell-Cattaneo \cite{Zakari93}) and specialize their dynamics
equations to the specific case of spherically symmetric perturbations of
neutron stars. In doing this, we extend the multi-component framework of
\citet{Gavassino21bulk} to include neutrino luminosity. The numerical
framework, implemented in the new code \texttt{hydro-bulk-1D} \cite{Camelio22code}, and the results
of our simulations, are presented in the companion paper \citep{Camelio22b}.
Our aim is not to make physical predictions regarding bulk viscosity in neutron stars but
to check if, in view of the mathematical duality between bulk-viscous fluids and reacting mixtures \citep{Gavassino21bulk},
the multi-component reacting fluid can be a numerically convenient alternative
to M\"uller-Israel-Stewart hydrodynamics for neutron star simulations.

The paper is organized as follows.
In Sec.~\ref{sec:intro-bulk} we give a simple example of bulk viscosity and we
outline different approaches to bulk viscosity in the literature.
In Sec.~\ref{sec:theory} we discuss in more detail the three different
approaches to bulk viscosity adopted in this paper.
In Sec.~\ref{sec:dual1} we discuss the relationship between these approaches and we extend the
M\"uller-Israel-Stewart theories to account for the energy loss during the
reactions.
In Sec.~\ref{sec:cs} we discuss the propagation speed of a signal in the
hydrodynamic equations and how it is related to the sound speed of the matter.
In Sec.~\ref{sec:microphysics} we describe our choice of the equation of state
and particle reaction rates.
In Sec.~\ref{sec:bulk-eqs} we specialize the equations to radial gauge,
polar slicing coordinates in spherical symmetry (i.e., Schwarzschild).
We draw our conclusions in Sec.~\ref{sec:conclusions}.
In Appendix~\ref{sec:cs-vs-cs} we derive some results
concerning the speed of sound in bulk viscous systems.
Finally, in Appendix~\ref{sec:signal-speed} we compute the characteristic
velocities of the Hiscock-Lindblom theory in non-linear regime.

We adopt the signature $(-,+,+,+)$ and, unless otherwise specified, we set $c=G=k_\mathrm{B}=1$.

\section{A brief introduction to bulk viscosity}
\label{sec:intro-bulk}

In this section, we give a brief introduction to bulk viscosity, including
a brief overview of the main approaches existing in the literature.

\subsection{A simple example of reaction-induced bulk viscosity}

\begin{figure*}
\centering
\begin{tikzpicture}[
   style-axes/.style={-{>[scale=1.5]}},
   style-thermo1/.style={postaction={decorate},decoration={markings,mark=at position 0.75 with {\arrow{>}}}},
   style-thermo2/.style={postaction={decorate},decoration={markings,mark=at position 0.75 with {\arrow{>}},mark=at position 0.25 with {\arrow{<}}}},
]

\begin{scope}[shift={(-1.5,-8.5)},rotate=90]
   \pgfmathsetseed{1}
   \draw[thick,gray] plot [only marks,mark=o,samples=75,mark size=1,domain=1:4.6] (\x,{rnd*2+1});
   \draw[thick] plot [only marks,mark=x,samples=50,mark size=2,domain=1:4.6] (\x,{rnd*2+1});
   \draw[very thick,fill=black!10!white] (0.6,0.6) -- (6,0.6) -- (6,1) -- (1,1) -- (1,3) -- (6,3) -- (6,3.4) -- (0.6,3.4) -- (0.6,0.6);
   \draw[very thick,fill=black!10!white] (6.5,1.8) -- (5,1.8) -- (5,1) --  (4.6,1) -- (4.6,3) -- (5,3) -- (5,2.2) -- (6.5,2.2);
   \draw[thick,<->] (5.25,2.5) -- (6.25,2.5);
   \node[anchor=south,align=center] at (6.5,2.5) {$\tau_\mathrm{pert}$};
   \node[anchor=south,align=center] at (-0.6,2) {$\tau_\mathrm{react}$};
   \draw[thick,gray] (-0.7,3) circle (.5mm);
   \draw[thick] (-0.7-0.05,1-0.05) -- (-0.7+0.05,1+0.05);
   \draw[thick] (-0.7-0.05,1+0.05) -- (-0.7+0.05,1-0.05);
   \draw[thick,<->] (-0.7,1.2) -- (-0.7,2.8);
\end{scope}

\node[anchor=center,align=center] at (1.5,0.2) {frozen\\($\tau_\mathrm{pert}\ll\tau_\mathrm{reac}$)};
\node[anchor=center,align=center] at (5.5,0.2) {viscous\\($\tau_\mathrm{pert}\approx\tau_\mathrm{reac}$)};
\node[anchor=center,align=center] at (9.5,0.2) {quasi-stationary\\($\tau_\mathrm{pert}\gg\tau_\mathrm{reac}$)};

\begin{scope}[shift={(0,-3.5)}]
   \draw[style-axes] (0,0) -- (0,3);
   \draw[style-axes] (0,0) -- (3,0);
   \node[anchor=north,align=center] at (1.5,0) {volume};
   \node[anchor=south,align=center,rotate=90] at (0,1.5) {pressure};
   \draw[fill] (0.7,2.3) circle (0.3mm) node[anchor=north east,align=center] {A};
   \draw[fill] (2.3,0.7) circle (0.3mm) node[anchor=south west,align=center] {B};
   \draw[style-thermo2] (2.3,0.7) arc (270:180:1.6);
\end{scope}

\begin{scope}[shift={(4,-3.5)}]
   \draw[style-axes] (0,0) -- (0,3);
   \draw[style-axes] (0,0) -- (3,0);
   \node[anchor=north,align=center] at (1.5,0) {volume};
   \node[anchor=south,align=center,rotate=90] at (0,1.5) {pressure};
   \draw[fill] (0.7,2.0) circle (0.3mm) node[anchor=north east,align=center] {A};
   \draw[fill] (2.3,0.7) circle (0.3mm) node[anchor=south west,align=center] {C};
   \draw[fill] (0.7,2.3) circle (0.3mm) node[anchor=south east,align=center] {A$^\prime$};
   \draw[style-thermo1] (0.7,2.0) .. controls (1,1) and (1.8,0.7) .. (2.3,0.7);
   \draw[style-thermo1] (2.3,0.7) .. controls (2,2) and (1,2.3) .. (0.7,2.3);
\end{scope}

\begin{scope}[shift={(8,-3.5)}]
   \draw[style-axes] (0,0) -- (0,3);
   \draw[style-axes] (0,0) -- (3,0);
   \node[anchor=north,align=center] at (1.5,0) {volume};
   \node[anchor=south,align=center,rotate=90] at (0,1.5) {pressure};
   \draw[fill] (0.7,2.3) circle (0.3mm) node[anchor=north east,align=center] {A};
   \draw[fill] (2.3,0.7) circle (0.3mm) node[anchor=south west,align=center] {D};
   \draw[style-thermo2] (2.3,0.7) arc (270:180:1.6);
\end{scope}

\begin{scope}[shift={(0,-7.0)}]
   \draw[style-axes] (0,0) -- (0,3);
   \draw[style-axes] (0,0) -- (3,0);
   \node[anchor=north,align=center] at (1.5,0) {volume};
   \node[anchor=south,align=center,rotate=90] at (0,1.5) {composition};
   \draw[fill] (0.7,1.5) circle (0.3mm) node[anchor=south east,align=center] {A};
   \draw[fill] (2.3,1.5) circle (0.3mm) node[anchor=south west,align=center] {B};
   \draw[style-thermo2] (0.7,1.5) -- (2.3,1.5);
\end{scope}

\begin{scope}[shift={(4,-7.0)}]
   \draw[style-axes] (0,0) -- (0,3);
   \draw[style-axes] (0,0) -- (3,0);
   \node[anchor=north,align=center] at (1.5,0) {volume};
   \node[anchor=south,align=center,rotate=90] at (0,1.5) {composition};
   \draw[fill] (0.7,0.7) circle (0.3mm) node[anchor=north east,align=center] {A};
   \draw[fill] (2.3,2.3) circle (0.3mm) node[anchor=south west,align=center] {C};
   \draw[fill] (0.7,1.0) circle (0.3mm) node[anchor=south east,align=center] {A$^\prime$};
   \draw[style-thermo1] (0.7,0.7) .. controls (1.4,0.7) and (2.3,1.5) .. (2.3,2.3);
   \draw[style-thermo1] (2.3,2.3) .. controls (1.5,2.3) and (0.7,1.7) .. (0.7,1.0);
\end{scope}

\begin{scope}[shift={(8,-7.0)}]
   \draw[style-axes] (0,0) -- (0,3);
   \draw[style-axes] (0,0) -- (3,0);
   \node[anchor=north,align=center] at (1.5,0) {volume};
   \node[anchor=south,align=center,rotate=90] at (0,1.5) {composition};
   \draw[fill] (0.7,0.7) circle (0.3mm) node[anchor=north east,align=center] {A};
   \draw[fill] (2.3,2.3) circle (0.3mm) node[anchor=south west,align=center] {D};
   \draw[style-thermo2] (0.7,0.7) arc (180:90:1.6);
\end{scope}

\begin{scope}[shift={(0,-10.5)}]
   \draw[style-axes] (0,0) -- (0,3);
   \draw[style-axes] (0,0) -- (3,0);
   \node[anchor=north,align=center] at (1.5,0) {entropy};
   \node[anchor=south,align=center,rotate=90] at (0,1.5) {temperature};
   \draw[fill] (1.5,2.3) circle (0.3mm) node[anchor=south west,align=center] {A};
   \draw[fill] (1.5,0.7) circle (0.3mm) node[anchor=north east,align=center] {B};
   \draw[style-thermo2] (1.5,0.7) -- (1.5,2.3);
\end{scope}

\begin{scope}[shift={(4,-10.5)}]
   \draw[style-axes] (0,0) -- (0,3);
   \draw[style-axes] (0,0) -- (3,0);
   \node[anchor=north,align=center] at (1.5,0) {entropy};
   \node[anchor=south,align=center,rotate=90] at (0,1.5) {temperature};
   \draw[fill] (0.8,1.7) circle (0.3mm) node[anchor=south east,align=center] {A};
   \draw[fill] (1.5,0.7) circle (0.3mm) node[anchor=north east,align=center] {C};
   \draw[fill] (2.3,2.3) circle (0.3mm) node[anchor=south west,align=center] {A$^\prime$};
   \draw[style-thermo1] (0.8,1.7) .. controls (0.8,1.2) and (1.0,0.7) .. (1.5,0.7);
   \draw[style-thermo1] (1.5,0.7) .. controls (2.0,0.7) and (2.3,1.5) .. (2.3,2.3);
\end{scope}

\begin{scope}[shift={(8,-10.5)}]
   \draw[style-axes] (0,0) -- (0,3);
   \draw[style-axes] (0,0) -- (3,0);
   \node[anchor=north,align=center] at (1.5,0) {entropy};
   \node[anchor=south,align=center,rotate=90] at (0,1.5) {temperature};
   \draw[fill] (1.5,2.3) circle (0.3mm) node[anchor=south west,align=center] {A};
   \draw[fill] (1.5,0.7) circle (0.3mm) node[anchor=north east,align=center] {D};
   \draw[style-thermo2] (1.5,0.7) -- (1.5,2.3);
\end{scope}
\end{tikzpicture}

\caption{
A mixture of two reactive particle species in an isolated container is a
minimal example of a multi-component system with dissipative response to
compression and expansion (see e.g.~Sec.~22.2 of \citet{MTW_book}), which gives
rise to bulk viscosity. The relative timescales of the perturbation $\tau_\mathrm{pert}$
and the reaction $\tau_\mathrm{reac}$ define the evolution regimes as discussed
in the text (see also \citep{haensel_frozen_2002,Gavassino21bulk}). With the label
`composition' we refer to, for example, the fraction of one particle species.
}
\label{fig:exe}
\end{figure*}
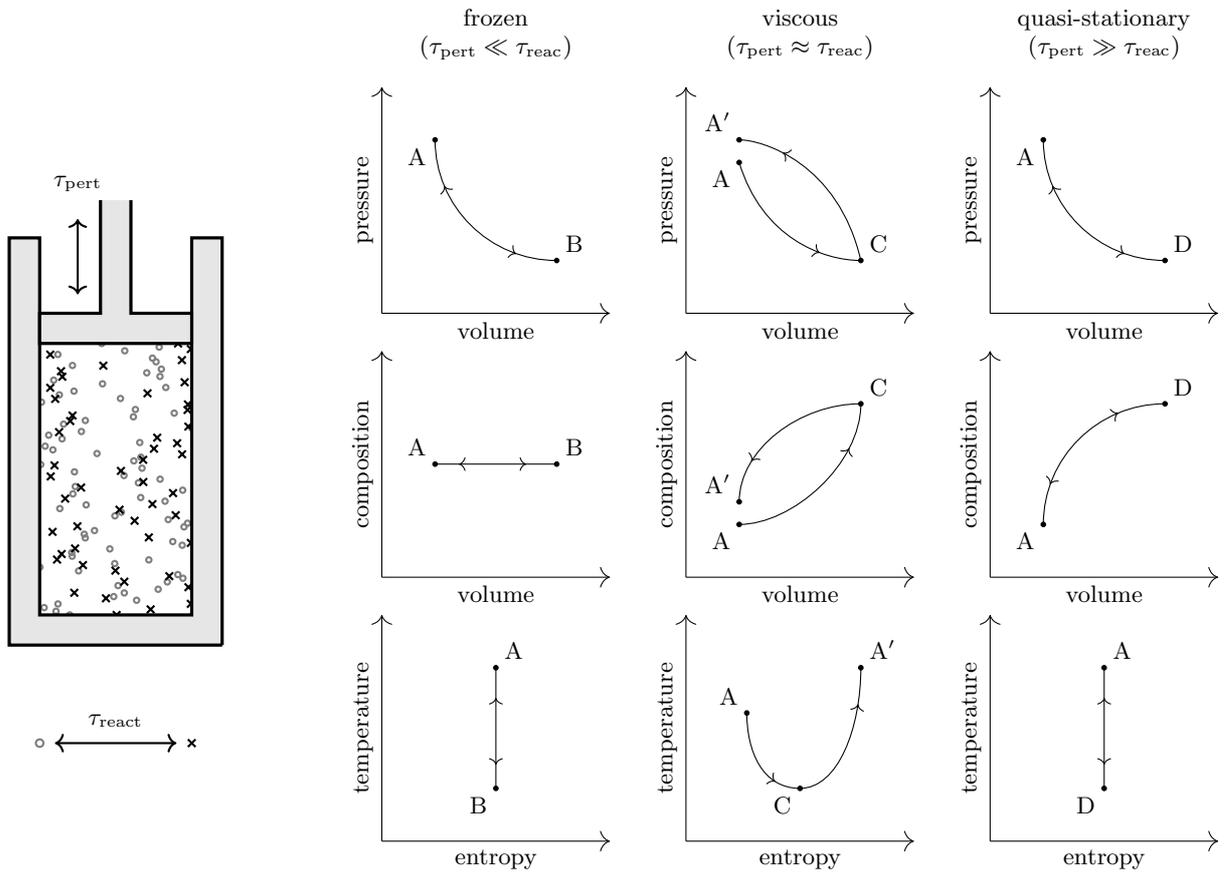

To illustrate how chemical reactions give rise to bulk viscosity, we
consider a thermally isolated system (see Fig.~\ref{fig:exe}) composed by a
mixture of two reacting particle species in the initial equilibrium state `A'.
When the system is driven out of equilibrium by, say, moving a piston, the reactions 
between the two particle species are imbalanced, with the consequence that chemical
transfusion tends to adjust the composition to a new equilibrium on a timescale
$\tau_\mathrm{reac}$, defined by the kinetics of the reactions.  
This system is a simplified model of what happens, locally, in a fluid
element of a neutron star, neglecting heat conduction and neutrino emission
(see e.g.~Sec.~22.2 of \citet{MTW_book}).  How the system reacts depends on how
$\tau_\mathrm{reac}$ compares with the timescale $\tau_\mathrm{pert}$ of the
volume perturbation: (i) if the perturbation is much faster than the reaction,
we are in the `frozen' regime \citep{haensel_frozen_2002}: the composition of the mixture is frozen and the perturbation is reversible. Since the system is thermally isolated
\cite{Landau5}, there is no heat exchange and the entropy is constant. (ii) If
the perturbation is much slower than the reaction, we are in the
`quasi-stationary' regime: the mixture `instantaneously' (when compared to the
perturbation timescale) adjusts itself, the
system evolves on a sequence of equilibrium states, the transformation is
reversible, and since there is no heat exchange, the entropy is constant. (iii)
If the perturbation and the reaction timescales are of the same order of
magnitude, we are in the `viscous' regime: the transfusion reactions change the
matter composition, but they do not have enough time to push the matter to a
new equilibrium.  This process is irreversible, and as a consequence the
entropy of the system grows even if there is no heat exchange
\cite{Jaynes1965}, which means that after a perturbation cycle the system is no more
in its initial state.

\subsection{Main approaches to bulk viscosity in the literature}

There are different approaches to bulk viscosity, and, more generally, to dissipation in relativity. 
We briefly mention only those that are relevant to the present paper.

The so-called first-order models (in the hydrodynamic gradients) are
relativistic versions of the Navier-Stokes equations, where the dissipative
fluxes (e.g., viscous stresses and heat current) are determined only by
first-order spatial gradients of the primary hydrodynamic variables of the
perfect fluid.  This approach, as originally formulated by \citet{Eckart40} and
\citet{landau6}, leads to superluminal signal propagation and nonphysical
instabilities \citep{Hiscock_Insatibility_first_order}.  In the context of
first-order gradient-expansion models, the causality and stability problems
have been only recently solved by \citet{Bemfica22}, by changing the definition
of temperature, chemical potential, and fluid velocity (see \citet{Kovtun19}).
However, like any model based on a gradient expansion, this approach (called
Bemfica-Disconzi-Noronha-Kovtun) is applicable only in the low-frequency limit.

M\"uller, Israel and Stewart \citep{Israel79, Israel_Stewart_1979} followed a
different approach. They adopted the prescription of extended irreversible
thermodynamics \citep{Jou_Extended, Frontiers_2021} of promoting the
dissipative fluxes to new dynamic degrees of freedom.  The resulting theory is
by construction causal \citep{Hiscock1983}, and, as a consequence
\cite{GavassinoSuperluminal2021}, stable
\citep{GavassinoLyapunov2020,GavassinoStabCarter2022}. Moreover, since the
expansion is performed near local equilibrium, and not for small
gradients\footnote{M\"uller-Israel-Stewart theories have been historically
confused with the more complicated `second-order' theories (in the hydrodynamic
gradients), because of similarities in the field equations
\cite{BaierRom2008}. However, strictly speaking, this is a misconception,
because \citet{Israel_Stewart_1979} performed a first-order expansion in
deviations from local equilibrium \cite{superbulk_PRD_2022}, and not a second
order expansion in the gradients. Failure to recognize this subtlety leads to
incorrect estimates of the relaxation timescale \cite{NoronhaOrigin2011}.},
M\"uller-Israel-Stewart theories are able (in some cases \cite{Heller2014}) to
accurately describe the dynamics of the slowest non-equilibrium degrees of
freedom \cite{Grozdanov2019,superbulk_PRD_2022}. This makes them applicable
also when $\tau_\mathrm{reac} \approx \tau_\mathrm{hydro}$, where
$\tau_\mathrm{hydro}$ is the timescale of the hydrodynamic process under
consideration. The same is not true for gradient-expansion theories, which are
built on the assumption $\tau_\mathrm{reac} \ll \tau_\mathrm{hydro}$
\cite{Kovtun19}. Therefore, given that our goal is precisely to explore the
regime $\tau_\mathrm{reac} \approx \tau_\mathrm{hydro}$, we cannot rely on the
Bemfica-Disconzi-Noronha-Kovtun theory \cite{Bemfica22, Kovtun19} and we must
follow the M\"uller-Israel-Stewart approach instead. 

  We consider two classes of M\"uller-Israel-Stewart theories: the
Hiscock-Lindblom theory \citep{Hiscock1983} and its linearization, the
Maxwell-Cattaneo \cite{Zakari93} theory\footnote{The linearization is performed with respect to
the additional dynamic degree of freedom called `bulk stress', and not on the
hydrodynamic gradients.}, which are both based on a perturbative expansion near local thermodynamic
equilibrium. Therefore, it is not surprising that M\"uller-Israel-Stewart
theories can still develop non-causal and unstable solutions when the
deviations from equilibrium become large \citep{Hiscock1983, Hiscock91,
Zakari93, Maartens95, GavassinoLyapunov2020}. Note that in all these frameworks, bulk viscosity does not necessarily arise from chemical
transfusion \cite{UdeyRadiation1982}, and indeed the M\"uller-Israel-Stewart and the Navier-Stokes models do not include any
chemical fraction.

\subsection{Mathematical duality between M\"uller-Israel-Stewart theories and
multi-component reacting fluids}

Recently, \citet{Gavassino21bulk} proposed a framework for bulk-viscous fluids
that builds on the formalism of \citet{carter1991} and naturally allows for
hyperbolic equations and causal solutions \citep{GavassinoStabCarter2022}, which are the
requirements needed for numerical applications and for thermodynamic stability \citep{Gavassino21stable,GavassinoSuperluminal2021}.
As shown in \citet{Gavassino21bulk}, when the quasi-equilibrium state of a fluid departing from full equilibrium can be
parametrized by a number of scalar variables, it can be mapped into a chemically
reacting mixture. In other words, these scalar variables can always be chosen in such
a way that they play the role of fictitious chemical fractions or, better, reaction
coordinates. This means that there is a mathematical duality between multi-component reacting fluids
mixtures and M\"uller-Israel-Stewart theories. 
The fact that this mapping is possible for a generic (possibly complex) fluid is not immediately obvious, for two reasons.  
First, chemical reactions are not necessary in order to have bulk viscosity: in general, a non-reacting fluid may have a non-zero bulk viscosity coefficient. 
Second, a bulk-viscous fluid modeled within the M\"uller-Israel-Stewart
framework has an energy-momentum tensor that departs from the multi-component fluid one, while the
model with fictitious chemical components of \citet{Gavassino21bulk} 
has an energy-momentum tensor that is formally the same as the one of a
multi-component fluid \citep{carter1991}.

In the special case in which only one scalar variable is sufficient to
parametrize the displacement from equilibrium, a near-equilibrium expansion of
the multi-component framework reproduces the M\"uller-Israel-Stewart hydrodynamics of a
purely bulk viscous fluid \citep{Gavassino21bulk}. Therefore, since Carter's
approach is not based on any near equilibrium expansion, in principle the
multi-component framework extends the M\"uller-Israel-Stewart one far from
equilibrium and to a generic number of reaction coordinates, possibly including
superfluid and heat conduction \citep{carter1991,superbulk_PRD_2022}, see the scheme in Fig.~1 in
\citet{superbulk_PRD_2022}.

\section{Three frameworks for bulk viscosity}
\label{sec:theory}

In this section we introduce the three frameworks for bulk viscous fluids considered in
this paper and numerically implemented in the companion paper
\citep{Camelio22b, Camelio22code}.

\subsection{Multi-component reacting fluid}
\label{ssec:mcpf}

As discussed in Sec.~\ref{sec:intro}, a fluid composed by two or more species
whose transfusion timescale is comparable with the hydrodynamic timescale
exhibits reaction-induced bulk viscosity, and in this case it is natural to
adopt the exact multi-component fluid approach of \citet{carter1991}.  
This means that the effect of reaction-induced bulk viscosity will be automatically implemented just by
keeping track of the change of chemical composition and by the consistent
implementation of the second law of thermodynamics \citep{Gavassino21bulk}.

The composition of a fluid with out-of-equilibrium reactions is not
determined by the other thermodynamic quantities, but has to be treated as a
set of independent hydrodynamic degrees of freedom. In this case of incomplete equilibrium, the first
law of thermodynamics is written in terms of an equation of state (EOS) that depends on the chemical fractions:
\begin{equation}
\label{eq:eos-th}
\mathrm du(\rho,s,\{Y_i\}_i)=\frac{p}{\rho^2}\mathrm d\rho+\frac{T}{m_\rn}\mathrm ds
-\sum_i\frac{\mathbb A^i}{m_\rn}\mathrm dY_i,
\end{equation}
where $\rho$ is the rest mass density, $u$ is the internal specific (per unit
mass) energy such that $\epsilon=(1+u)\rho$ is the total (including rest mass)
energy density, $p$ is the pressure, $m_\rn$ the neutron mass, $T$ the
temperature, $s$ the entropy per baryon, $Y_i$ the number fraction (with
respect to the total baryon number) of the independent particle species $i$,
and $\mathbb A^i$ is the affinity of particle $i$.
The system reaches complete thermodynamic equilibrium when $\mathbb A^i=0$ for
all the independent species $i$ in the EOS $u(\rho,s,\{Y_i\}_i)$.

Since we are interested in neutron stars, we consider a multi-component general
relativistic fluid whose species react and emit neutrinos. To simplify the
discussion, we assume that neutrinos are emitted isotropically in the fluid
frame and immediately leave the star\footnote{For the complete equations
including the neutrino momentum, see Appendix~A of \citet{OConnor10}.
Note that our luminosity $\mathcal Q$ has the opposite sign of their `net heat' $Q$
\citep{OConnor10}.}.  The
hydrodynamic equations for such a system are determined by the continuity
equation for each species and by the energy and momentum conservation equations
\cite{OConnor10}:
\begin{align}
\label{eq:continuity}
\nabla_\mu \big( \rho u^\mu \big) ={}&0,\\
\label{eq:pf-conservation}
\nabla_\mu\big( T^{\mu \nu}\big)={}&-\mathcal Q u^\nu,\\
\label{eq:pf-chemical}
\nabla_\mu \big( \rho Y_i u^\mu \big)={}&m_\rn\mathcal R_i,
\end{align}
where $u^\nu$ is the four-velocity of the matter, $\mathcal Q=\sum_i\mathcal Q_j$ is the total
luminosity (i.e., the rate of energy loss of all reactions; $\mathcal Q_j$ is the luminosity
of reaction $j$), $\mathcal R_i$ is the number reaction rate of particle
species $i$.
Finally, $T^{\mu\nu}$ is the stress-energy tensor of a multi-component fluid,
\begin{equation}
\label{eq:pf-tmunu}
T^{\mu\nu}=(\epsilon + p)u^\mu u^\nu + pg^{\mu\nu},
\end{equation}
where $g^{\mu\nu}$ is the metric. Although this multi-component system is dissipative 
(see, e.g., Sec. III of \citep{Gavassino21bulk}), its stress-energy tensor \eqref{eq:pf-tmunu} retains 
the form it would have in the absence of dissipation 
(namely, when $\mathcal{R}_i=0$, $\mathcal Q =0$).

\subsection{Hiscock-Lindblom}
\label{ssec:is}

Bulk viscosity can also be accounted for if we astray the composition from the
description of the matter.  In the Hiscock-Lindblom theory of bulk viscosity,
the hydrodynamic equations are (Eqs.~(18) and (19) of \citet{Hiscock1983}):
\begin{align}
\label{eq:bv-continuity}
\nabla_\mu \big( \rho u^\mu \big) ={}&0,\\
\label{eq:bv-conservation}
\nabla_\mu\big( T^{\mu \nu}\big)={}&0,\\
\label{eq:bv-entropy}
\nabla_\mu\big(\rho s u^\mu\big)={}&\frac{m_\rn\Pi^2}{\zeta T^\req(\rho,\epsilon)},
\end{align}
where $\Pi$ is the bulk viscous stress, $\zeta$ the viscous coefficient, and
$s$ (the entropy per baryon of the out-of-equilibrium system) is now expanded to second order in~$\Pi$:
\begin{align}
\label{eq:bv-s}
s={}&s^\req(\rho,\epsilon) - \frac{\chi\Pi^2}{2nT^\req(\rho,\epsilon)},\\
\label{eq:bv-tau}
\tau={}&\zeta\chi,
\end{align}
where $n=\rho/m_\rn$ is the baryon number density, $\chi>0$ is the bulk viscous
parameter, $\tau$ is the bulk viscous timescale, and the `eq' superscript
refers to the quantities at equilibrium for given $\epsilon$ and
$\rho$,\footnote{\label{fnt:1} We work in the Eckart frame \cite{Kovtun19},
which fixes the fiducial local-equilibrium state by imposing $\rho^\req:=\rho$
and $\epsilon^\req:=\epsilon$. In principle, one could make a different choice;
for example, we could impose $\rho^\req:=\rho$ and $s^\req:=s$, in which case
a correction $\propto \Pi^2$ would be transferred from the right-hand side of
\eqref{eq:bv-s} to that of \eqref{eq:bv-tmunu}, as in \citet{Gavassino21bulk}. }
for which $\{Y_i=Y_i^\req\}_i$.
The bulk viscous stress-energy tensor $T^{\mu\nu}$ can be obtained from the
multi-component one in Eq.~\eqref{eq:pf-tmunu} with the decomposition $p= p^\mathrm{eq}(\rho,\epsilon) + \Pi$:
\begin{equation}
\label{eq:bv-tmunu}
T^{\mu\nu}=\big(\epsilon + p^\req(\rho,\epsilon) + \Pi\big)u^\mu u^\nu
+ \big(p^\req(\rho,\epsilon) + \Pi\big)g^{\mu\nu}.
\end{equation}

Eq.~\eqref{eq:bv-entropy} for the evolution of the entropy is
equivalent to (cf.~Eq.~(21) of \citet{Hiscock1983}):
\begin{equation}
\label{eq:telegraph-bulk}
\Pi=-\zeta\left[\nabla_\mu u^\mu +
\chi u^\mu\nabla_\mu\Pi+\frac\Pi2 T^\req\nabla_\mu\left(\frac{\chi
u^\mu}{T^\req}\right)\right],
\end{equation}
which is in the so-called `telegraph-type' form and is the one used in many
implementations, after rewriting it in a flux-conservative form:
\begin{multline}
\label{eq:hiscock-lindblom}
\nabla_\mu(\Pi u^\mu)=-\frac\Pi\tau -\left(\frac1\chi-\frac\Pi2\right)\nabla_\mu u^\mu\\
-\frac\Pi2 u^\mu\nabla_\mu\left(\log\frac{\chi}{T^\req}\right).
\end{multline}

\subsection{Maxwell-Cattaneo}
\label{ssec:mc}

The linearization in the bulk stress $\Pi$ of the Hiscock-Lindblom theory is obtained
by dropping the last term of Eq.~\eqref{eq:telegraph-bulk}:
\begin{equation}
\label{eq:max-cat}
\Pi=-\zeta\nabla_\mu u^\mu - \tau u^\mu\nabla_\mu\Pi.
\end{equation}
Eq.~\eqref{eq:max-cat} is known as Maxwell-Cattaneo equation,
and can be put in a flux-conservative form together with the other
hydrodynamic equations:
\begin{align}
\label{eq:mc-continuity}
\nabla_\mu \big( \rho u^\mu \big) ={}&0,\\
\label{eq:mc-conservation}
\nabla_\mu\big( T^{\mu \nu}\big)={}&0,\\
\label{eq:maxwell-cattaneo}
\nabla_\mu(\Pi u^\mu)={}&-\frac\Pi\tau -\left(\frac1\chi-\Pi\right)\nabla_\mu u^\mu.
\end{align}
Note that, while Maxwell-Cattaneo is linear in the bulk stress $\Pi$,
it is not linear in general since at first order $\nabla_\mu u^\mu\propto \Pi$.

Note also that these formulations of Hiscock-Lindblom and Maxwell-Cattaneo do
not take into account that neutrinos are leaving the star carrying out energy.
We address this issue in Sec.~\ref{ssec:q}.

\section{Duality between M\"uller-Israel-Stewart and the multi-component fluid}
\label{sec:dual1}

In this section we show how to map the multi-component fluid model introduced in
Sec.~\ref{ssec:mcpf}, that is based on the approach of \citet{carter1991} and
does not explicitly contain the bulk stress $\Pi$, into the M\"uller-Israel-Stewart
model (either Hiscock-Lindblom or Maxwell-Cattaneo), and viceversa, for a purely bulk-viscous
fluid (i.e., in the absence of heat conduction, superfluidity and shear
viscosity).  Further details on the mathematical aspects of this
mapping are given in \citet{Gavassino21bulk}, while the extension of the
mapping to superfluid and heat conducting fluids is developed in
\citet{superbulk_PRD_2022}.

\subsection{Mapping the multi-component fluid into the M\"uller-Israel-Stewart model}
\label{ssec:carter2is}

Given a multi-component model with a non-equilibrium EOS and reaction rates $\mathcal{R}_i$,
it is possible to derive the equivalent bulk viscous parameters that appear in its dual
M\"uller-Israel-Stewart theories, see Eqs.~(84) and (27) of \citet{Gavassino21bulk}:
\begin{align}
\label{eq:def-zeta}
\zeta={}&n^4\Xi^{ab}\frac{\partial Y_a^\req}{\partial n}\bigg|_s\frac{\partial Y_b^\req}{\partial n}\bigg|_s,\\
\label{eq:def-xi}
\Xi_{ab}={}&\left.\frac{\partial \mathcal R_a(\{\mathbb A^j=0\}_{\forall j})}{\partial \mathbb A^b}\right|_{\rho,s,\{\mathbb A^i\}_{i\neq b}},
\end{align}
where the indices $a,b$, that label the fluid components are raised and lowered by matrix inversion ($\Xi_{ab}\Xi^{bc}=\delta^c_a$), and are manipulated according to Einstein's sum convention.
If the system can be described by a single out-of-equilibrium fraction, we can
introduce a reaction timescale, see Eq.~(113) of \citet{Gavassino21bulk},
\begin{equation}
\label{eq:tau-from-tracking}
\tau=-\frac{n}{\Xi}\left.\frac{\partial Y(\mathbb A=0)}{\partial \mathbb A}\right|_{\rho,s},
\end{equation}
where we dropped the species index because there is only one independent reaction coordinate ($\Xi\equiv\Xi_{11}$).
Moreover, we can generalize the bulk-viscous timescale to more than one species
out-of-equilibrium by requiring that the speed of sound of the
multi-component fluid coincide with that of the M\"uller-Israel-Stewart
theory [see Sec.~\ref{sec:cs} and Appendix~\ref{ssec:is-vs-mcf}]:
\begin{equation}
\label{eq:tau-multispecies}
\tau = \dfrac{n \Xi^{ab} \,  \dfrac{\partial Y_a^\req}{\partial \rho} \bigg|_s
\,  \dfrac{\partial Y_b^\req}{\partial \rho} \bigg|_s}{ m_\rn \dfrac{\partial^2
u}{\partial Y_c \partial Y_d} \bigg|_{\rho,s} \,  \dfrac{\partial
Y_c^\req}{\partial \rho} \bigg|_s \, \dfrac{\partial Y_d^\req}{\partial \rho}
\bigg|_s} \, ,
\end{equation}
where we recall that $u=\epsilon/\rho-1$ is the specific internal energy.
For a single chemical fraction, Eq. \eqref{eq:tau-multispecies} reduces to \eqref{eq:tau-from-tracking}. 

\subsection{Mapping M\"uller-Israel-Stewart into the multi-component fluid}
\label{ssec:is2carter}

As anticipated in Sec \ref{sec:intro}, both thermodynamic and chemical
processes contributing to bulk viscosity can be described in terms of a set of
scalar variables (reaction coordinates) that parametrize the displacement of
the substance from local thermodynamic equilibrium.  This results in a
multi-component fluid with comoving and reacting chemical species, that can be
interpreted as a universal model for purely bulk-viscous (no shear viscosity
and no heat conduction) fluids \citep{Gavassino21bulk, Frontiers_2021}.
Therefore, given a bulk-viscous M\"uller-Israel-Stewart fluid, it is possible to find an
equivalent multi-component description and viceversa.

In Sec.~\ref{ssec:carter2is} we have seen that it is straightforward to map the
multi-component model into the M\"uller-Israel-Stewart one. However, the
inverse mapping is more subtle.  In fact, the M\"uller-Israel-Stewart
models are defined by the parameters $\zeta(\rho,s)$ and
$\tau(\rho,s)$, but these are not enough to uniquely define a multi-component
fluid. Therefore, for a  given M\"uller-Israel-Stewart model it is possible to build a
whole family of multi-component fluids, that are all equivalent in the limit of small perturbations around equilibrium.
This is due to the fact that the M\"uller-Israel-Stewart framework is a
perturbative one, based on an expansion near equilibrium. Therefore, while a
multi-component fluid always gives a well-defined
M\"uller-Israel-Stewart model, the inverse process is similar to try to recreate a
function from its truncated Taylor series: the result cannot be unique.
Moreover, the reaction coordinate (i.e., the fictitious particle abundance
$Y$) must behave as a frozen variable in the high-frequency limit
\cite{Gavassino21bulk}.

As an example, we will show in this section that the two-component fluid defined by:
\begin{align}
u(\rho,s,Y)={}&u^\req(\rho,s) + \frac{\zeta}{2m_\rn n\tau}\big(Y-\log n\big)^2,\\
\mathcal R(\rho,s,Y)={}&-\frac{n}{\tau}\big(Y-\log n\big),
\end{align}
is equivalent to the Maxwell-Cattaneo model (which in turn is equivalent at
first order to the Hiscock-Lindblom theory). First, we obtain the affinity and
the equilibrium particle fraction:
\begin{align}
\mathbb A(\rho,s,Y)={}&-m_\rn\frac{\partial u}{\partial Y} \bigg|_{\rho,s}= -\frac{\zeta}{n\tau}\big(Y-\log n\big),\\
Y^\req(\rho,s)={}&Y(\rho,s,\mathbb A=0)=\log n \, .
\end{align}
We then verify, using Eqs.~\eqref{eq:def-zeta}--\eqref{eq:tau-from-tracking},
that the reacting multi-component model is equivalent to the M\"uller-Israel-Stewart (whose quantities
are identified with the `MIS' subscript) one:
\begin{align}
\Xi={}&\frac{\partial \mathcal R}{\partial \mathbb A}=\frac{n^2}{\zeta},\\
\zeta_{\text{MIS}}={}& \frac{n^4}{\Xi}\left(\frac{\partial Y^\req}{\partial n}\right)^2=\zeta,\\
\tau_{\text{MIS}}={}& -\frac{n}{\Xi}\frac{\partial Y}{\partial \mathbb A}=\tau.
\end{align}
From Eq.~\eqref{eq:a2pi} we have that
\begin{align}
\Pi={}& n\mathbb A,\\
Y={}& -\frac{\tau}{\zeta}\Pi + \log n,
\end{align}
from which we obtain ($\nabla_u= u^\mu\nabla_\mu$):
\begin{equation}
\label{eq:doty}
\nabla_u Y= -\nabla_u{\left(\frac{\tau}{\zeta}\Pi\right)} + \frac{\nabla_u n}{n}\simeq
-\frac{\tau}{\zeta}\nabla_u\Pi + \frac{\nabla_u n}{n},
\end{equation}
where in the last step we kept the first order contributions in the deviations from equilibrium.
From Eqs.~\eqref{eq:continuity} and \eqref{eq:pf-chemical} we have that
\begin{equation}
\label{eq:ndoty}
nu^\mu\nabla_\mu Y=\nabla_\mu(nYu^\mu)
= \mathcal R= \Xi \mathbb A=\frac{n\Pi}{\zeta},
\end{equation}
and inserting Eq.~\eqref{eq:doty} into~\eqref{eq:ndoty} and using $u^\mu\nabla_\mu n=- n\nabla_\mu u^\mu$
[where we used again Eq.~\eqref{eq:continuity}] we obtain the Maxwell-Cattaneo equation~\eqref{eq:max-cat}.

\subsection{A simple application of the mapping: extending M\"uller-Israel-Stewart to account for radiative losses}
\label{ssec:q}

A first advantage of the mapping outlined in Sec.~\ref{ssec:carter2is} is that it
allows us to consistently include the effects of neutrino emission in the
M\"uller-Israel-Stewart theories.

In general, the energy and momentum conservation equation~\eqref{eq:pf-conservation} 
has a contribution accounting for the energy
and momentum lost by the fluid due to the reactions (e.g., neutrinos escaping
from the star).  However, the original M\"uller-Israel-Stewart theory does not have this
contribution, cf.~Eqs.~\eqref{eq:pf-conservation},
\eqref{eq:bv-conservation}, and \eqref{eq:mc-conservation}.  Realizing how the
M\"uller-Israel-Stewart theory can be derived from the multi-component reacting fluid,
and in particular that at first order the bulk stress is given by
(cf.~Eq.~(65) of \citet{Gavassino21bulk}):
\begin{equation}
\label{eq:a2pi}
\Pi=\mathbb A^i\left.\frac{\partial Y_i^\req}{\partial n}\right|_{s}n^2,
\end{equation}
we can approximately account for the energy loss by expanding the luminosity $\mathcal Q$
around equilibrium:
\begin{align}
\label{eq:q}
\mathcal Q_\mathrm{bv}(\rho,s,\Pi)={}&\mathcal Q^\req(\rho,s^\req)
+\frac{\partial \mathcal Q}{\partial \Pi}\Pi + \mathcal O(\Pi^2),\\
\label{eq:dqdpi}
\frac{\partial \mathcal Q}{\partial \Pi}={}&
\frac1{n^2}\left(\left.\frac{\partial Y^\req}{\partial n}\right|_s\right)^{-1}
\left.\frac{\partial \mathcal Q(\mathbb A=0)}{\partial \mathbb A}\right|_{\rho,s}.
\end{align}
where the subscript `bv' means `bulk viscous' and we dropped the species index because we consider for simplicity
only one independent species. Note that in Eq. \eqref{eq:q} we could replace
$s$ with $s^\req$ because $s-s^\req$ is of second order in $\Pi$,
cf.~Eq.~\eqref{eq:bv-s}.  Finally, Eqs.~\eqref{eq:bv-conservation} and
\eqref{eq:mc-conservation} become:
\begin{equation}
\label{eq:conservation-q}
\nabla_\mu\big( T^{\mu \nu}\big)=-\mathcal Q_\mathrm{bv} u^\nu.
\end{equation}

\section{Propagation speed of a signal and sound speed of the matter}
\label{sec:cs}

In a hydrodynamic code, the propagation speed of a signal is used to compute
the timestep from the Levy-Friedrichs-Courant condition and to solve the
Riemann problem, and is equal to the maximal characteristic speed of the
hydrodynamic equations.  In a hydrodynamic system, the speed of
sound enters in the definition of the propagation speed of a signal and is
therefore needed to evolve the equations.

The definition of `speed of sound' of a reacting fluid depends on the frequency
of the sound wave.
If the period of the wave is much shorter than the reaction timescale ($\tau_\mathrm{pert}\ll\tau_\mathrm{react}$),
we are in the `ultraviolet' limit, namely the composition is frozen and
\begin{equation}
\label{eq:cs2-uv}
c_\mathrm{s,uv}^2(\rho,s,\{Y_i\}_i)=\left.\frac{\partial p}{\partial \epsilon}\right|_{s,\{Y_i\}_i}.
\end{equation}
The sound speed in Eq.~\eqref{eq:cs2-uv} is used in hydrodynamic codes that
evolve a multi-component fluid, because the timestep must be short enough to
resolve the dynamics, which is influenced by the particle fractions.

On the other hand, if the wave period is much longer than the reaction timescale
($\tau_\mathrm{pert}\gg\tau_\mathrm{react}$), we are in the `infrared'
limit, that means that the fluid is always at chemical equilibrium and 
\begin{equation}
\label{eq:cs2-ir}
c_\mathrm{s,ir}^2(\rho,s)=\left.\frac{\partial p(\{\mathbb{A}^i=0\}_i)}{\partial \epsilon}\right|_{s,\{\mathbb{A}^i\}_i}.
\end{equation}

In the limit in which the EOS contains only one independent species we recover
the hydrodynamics of a perfect fluid: in this case, Eq.~\eqref{eq:cs2-ir}
coincides with Eq.~\eqref{eq:cs2-uv}, because the thermodynamic quantities do
not depend on the particle fractions.
In general, however, if the equilibrium state is stable against fluctuations,
one can prove that (see Appendix~\ref{sec:uv-vs-ir}):
\begin{equation}
\label{eq:csuv-geq-csir}
c_\mathrm{s,uv}^{\req } \geq c_\mathrm{s,ir}.
\end{equation}

We can also define an infrared and an ultraviolet speed of sound
for the M\"uller-Israel-Stewart theories.
In the infrared limit, the gradients are small, hence the viscous stresses become
negligible and the fluid behaves as a perfect fluid:
\begin{equation}
\label{eq:csir-bulk}
c_{s,\mathrm{ir}}^2  = \dfrac{\partial p^\req}{\partial \epsilon}\bigg|_{s^\req}.
\end{equation}
In the ultraviolet limit, perturbations propagate along the characteristics
of the full M\"uller-Israel-Stewart theory.
In the Maxwell-Cattaneo case, the speed of sound is \cite{Bemfica19}:
\begin{equation}
\label{eq:cs-bemfica}
c_{s,\mathrm{uv}}^2  = \frac{\partial p^\req}{\partial \epsilon}\bigg|_\rho
+\frac{1}{\epsilon+p^\req+\Pi}\left(\rho\left.\frac{\partial p^\req}{\partial \rho}\right|_\epsilon
+ \frac{1}{\chi}\right),
\end{equation}
while in the Hiscock-Lindblom case, the speed of sound is given by Eq.~\eqref{eq:hl-propagation-speed}.
Close to equilibrium, we have in both cases that:
\begin{equation}
\label{eq:effective-cs2}
c_{s,\mathrm{uv}}^{\req \, 2}= c_\mathrm{s,ir}^2 + \frac{1}{(\epsilon+p^\req)\chi},
\end{equation}
and we recover the thermodynamic inequality \eqref{eq:csuv-geq-csir}.
We remark that stability and causality require that the last term of Eq.~\eqref{eq:effective-cs2}
does not diverge for vanishing rest mass density (Sec.~III~(b) of \citet{Hiscock1983}).

It is interesting to note that, since the ultraviolet speed of sound is the
actual signal propagation speed of the field equations, causality only demands that
$c_{s,\mathrm{uv}}^{2}<1$ and \textit{not} that $c_{s,\mathrm{ir}}^{2}<1$.
This implies that a viscous fluid may be consistent with the principle of
causality even if the infrared speed of sound is superluminal. However, such a
fluid would be thermodynamically unstable, and Eq.~\eqref{eq:csuv-geq-csir}
would not hold. This result is a generalization of the Bludman-Ruderman theorem
\cite{BludmanIstab1970} to reacting mixtures.

\section{Microphysics}
\label{sec:microphysics}

In order to close the hydrodynamic equations, it is necessary to introduce an
EOS.  Moreover, since bulk viscosity is due to reactions, we also need the
reaction rates $\mathcal R_i$ and $\mathcal Q_i$.  The most accurate way to do
this is by using the results of microphysics calculations in form of tables
\citep{Typel15, Oertel17}.  However, the focus of this paper and its companion
\citep{Camelio22b} is to compare different bulk viscous frameworks and not to
obtain accurate astrophysical predictions.  For this reason, we opted to use
simple, analytic, but at the same time realistic EOS and reaction rates, that
allow us to study bulk viscosity without the additional complications of table
interpolation.

\subsection{Equation of state}
\label{ssec:eos}

We consider a neutrinoless fluid of protons `p', neutrons `n', electrons `e', and muons `$\mu$',
which describe a mature (i.e., cold) neutron star, where neutrinos are not trapped.
We assume that the fluid undergoes only direct beta reactions (direct Urca):
\begin{align}
\label{eq:beta-minus-e}
\beta^-_\mathrm{e}:  {}&\quad \mathrm n \to \mathrm p + \mathrm e^- + \bar\nu_\mathrm{e},\\
\label{eq:beta-plus-e}
\beta^+_\mathrm{e}:  {}&\quad \mathrm p + \mathrm e^- \to \mathrm n + \nu_\mathrm{e},\\
\label{eq:beta-minus-mu}
\beta^-_\mu:{}&\quad \mathrm n \to \mathrm p + \mu^- + \bar\nu_\mu,\\
\label{eq:beta-plus-mu}
\beta^+_\mu:{}&\quad \mathrm p + \mu^- \to \mathrm n + \nu_\mu.
\end{align}
Since baryons are conserved ($\mathrm dY_\mathrm n=-\mathrm dY_\mathrm p$),
the system is charge-neutral ($\mathrm dY_\mathrm
p=\mathrm dY_\mathrm e+\mathrm dY_\mu$),
and neutrinos immediately leave the star ($Y_\nu=0$), then
there are only 2 independent chemical fractions, e.g.~electrons and muons, and
the first law of thermodynamics is:
\begin{align}
\label{eq:1st-law}
\mathrm du={}&\frac p{\rho^2}\mathrm d\rho + \frac T{m_\rn}\mathrm ds
- \frac{\mathbb A^\re}{m_\rn} \mathrm dY_\re
- \frac{\mathbb A^\mu}{m_\rn} \mathrm dY_\mu,\\
\label{eq:affinity}
\mathbb A^i={}& \mu_\mathrm{n} - \mu_\mathrm{p} - \mu_i.
\end{align}

Our equation of state is defined by:
\begin{align}
\label{eq:eos}
u={}& k_0\rho+ k_\rth s^2\rho^{\Gamma_\rth-1}
+ k_\re\Delta Y_\re^2 + k_\mu\Delta Y_\mu^2,\\
\Delta Y_i={}&{}Y_i - Y_i^0 \frac{\rho}{\rho_\rn},
\end{align}
where $u=u(\rho,s,Y_\re,Y_\mu)$,
$\Gamma_\rth$ is the thermal polytropic exponent, $k_0,k_\rth,k_\re,k_\mu>0$
are (positive) polytropic parameters, and $Y_\re^0,Y_\mu^0$ are the equilibrium values
of the particle fractions at saturation density.
We note that this EOS is an extension of a $\Gamma=2$ polytropic EOS (the
thermal component was introduced in \citet{Camelio19}), and that we are
assuming that the electron and muon polytropic exponents are $\Gamma_\re=\Gamma_\mu=1$.
The reason of these choices is that in this way the pressure and the speed of
sound are always positive\footnote{Strictly speaking, $p>0$ is not a
fundamental thermodynamic requirement: negative pressures are physically
possible in many interacting systems, also in the liquid state
\citep{Imre1998}.  However, we assume positive pressure to avoid possible
complications in the numerical implementation \cite{Camelio22code} presented in
\citet{Camelio22b}.} 
if $Y_i\in[0,1]$ and if
\begin{equation}
k_0>2\frac{{k_\re Y_\re^0 + k_\mu Y_\mu^0}}{\rho_\rn}.
\end{equation}

Our EOS has the advantage of being simple and analytic, but at the
same time it reproduces the qualitative features of a more realistic EOS. 
In particular, the cold EOS at equilibrium is a $\Gamma=2$ polytrope, which is a
common choice for testing new codes, and the temperature
goes to zero as the entropy goes to zero \cite{landau6}.
Moreover, the equilibrium fraction is analytical and given by:
\begin{equation}
Y^\req_i(\rho)=Y^0_i\frac{\rho}{\rho_\rn}.
\end{equation}
Note that, in order to have bulk viscosity,
$Y^\req_i$ must depend on $\rho$, otherwise $\zeta$ and $\Pi$ would vanish
identically, cf.~Eq. \eqref{eq:def-zeta} and \eqref{eq:a2pi}.

The other thermodynamic quantities can be derived
from the EOS [Eq.~\eqref{eq:eos}] and the first law of thermodynamics
[Eq.~\eqref{eq:1st-law}]:
\begin{align}
p={}&k_0\rho^2+ (\Gamma_\rth-1)k_\rth s^2\rho^{\Gamma_\rth} \notag\\
&{}- 2\frac{\rho^2}{\rho_\rn}\sum_i k_i Y_i^0 \Delta Y_i,\\
T={}& 2m_\rn k_\rth s \rho^{\Gamma_\rth-1},\\
\mathbb A^i={}&{}-2m_\rn k_i\Delta Y_i.
\end{align}
Moreover, the ultraviolet and infrared sound speeds defined in Sec.~\ref{sec:cs} are:
\begin{align}
c_\mathrm{s,uv}^2={}&\frac{\left.\frac{\partial p^\req}{\partial\rho}\right|_s
-2\frac{\rho}{\rho_\rn}\sum_ik_iY_i^0\left(2\Delta Y_i- Y_i^0\frac{\rho}{\rho_\rn}\right)}
{\left.\frac{\partial\epsilon^\req}{\partial\rho}\right|_s
+\sum_i k_i \Delta Y_i\left(\Delta Y_i - 2 Y_i^0 \frac{\rho}{\rho_\rn}\right)},\\
c_\mathrm{s,ir}^2={}&\left.\frac{\partial p^\req}{\partial\rho}\right|_s\div
\left.\frac{\partial\epsilon^\req}{\partial\rho}\right|_s,\\
\left.\frac{\partial p^\req}{\partial\rho}\right|_s={}&{}2k_0\rho+\Gamma_\rth(\Gamma_\rth-1)k_\rth s^2\rho^{\Gamma_\rth-1},\\
\left.\frac{\partial\epsilon^\req}{\partial\rho}\right|_s={}&{}1+2k_0\rho+\Gamma_\rth k_\rth s^2\rho^{\Gamma_\rth-1}.
\end{align}
Note that, as pointed out in Sec.~\ref{sec:cs}, $c_\mathrm{s,uv}^\req \geq c_\mathrm{s,ir}$.

\subsection{Reaction rates}
\label{ssec:rates}

In this section we use cgs units, assume that the particle species are
Fermi liquids, that the matter is strongly degenerate, that $\mathbb A^i \ll
\mu_\rn,\mu_\mathrm p,\mu_i$ ($i=\{\re,\mu\}$), approximate the nucleon
effective masses to their bare ones, and neglect the effects of the
medium on the weak interactions \cite{Haensel92}. With these
approximations, the number $\mathcal R_i^-$ and energy $\mathcal Q_i^-$ rates
of the neutron decay reactions $\beta_i^-$ [i.e.~direct Urca reactions, see
Eqs.~\eqref{eq:beta-minus-e} and \eqref{eq:beta-minus-mu}] are (Eqs.~(6)--(9)
of \citet{Haensel92}):
\begin{align}
\mathcal R_i^-={}&\frac{8.86\times 10^{31}}{\unit{cm^{3}\,s}}\sqrt[3]{\frac{Y_i\rho}{\rho_n}}
\left(\frac{T}{\unit[10^9]{K}}\right)^5 G(x_i)
,\\
\label{eq:g-def}
G(x)={}&\int_0^\infty y^2\frac{\pi^2 + (y - x)^2}{1 + \exp(y-x)}\mathrm dy,\\
\mathcal Q_i^-(x)={}&\frac{1.22\times 10^{25}}{\unit{erg^{-1}\,cm^{3}\,s}}\sqrt[3]{\frac{Y_i\rho}{\rho_n}}
\left(\frac{T}{\unit[10^9]{K}}\right)^6 F(x_i),\\
\label{eq:f-def}
F(x)={}&\int_0^\infty y^3\frac{\pi^2+(y-x)^2}{1+\exp(y-x)}\mathrm dy,\\
x_i={}&\frac{\mathbb A^i}{k_\mathrm{B}T}.
\end{align}
In our case, the number $\mathcal R_i^+$ and energy $\mathcal Q_i^+$ rates of the lepton capture reactions $\beta^+_i$
[Eqs.~\eqref{eq:beta-plus-e} and \eqref{eq:beta-plus-mu}] and the total
number $\mathcal R_i$ and energy $\mathcal Q_i$ rates of the direct beta reactions are \cite{Haensel92}:
\begin{align}
\mathcal R_i^+(x_i)={}& \mathcal R_i^-(-x_i),\\
\mathcal Q_i^+(x_i)={}&\mathcal Q_i^-(-x_i),\\
\label{eq:tot-nrate}
\mathcal R_i={}& \mathcal R_i^- - \mathcal R_i^+,\\
\label{eq:tot-erate}
\mathcal Q_i={}&\mathcal Q_i^- + \mathcal Q_i^+.
\end{align}

From the definition of polylogarithm (Eq.~(A.3.8.2) of \citet{Lewin81}):
\begin{equation}
\label{eq:polylog-def}
\mathrm{Li}_n(z)=\frac1{\Gamma(n)}\int_0^\infty \frac{zy^{n-1}}{\mathrm e^y-z}\mathrm dy,
\quad \mathrm{Re}(n)>0,
\end{equation}
and assuming that $z\in \mathbb R^-$ and $n\in \mathbb N^+$, we have
\begin{equation}
\label{eq:polylog}
\int_0^\infty \frac{y^{n-1}}{1+\exp(y-x)}\mathrm dy=-\mathrm{Li}_n(-\mathrm e^x)(n-1)!,
\end{equation}
where $x= \log(-z)$.
Substituting Eq.~\eqref{eq:polylog} in $\mathcal R_i$ and $\mathcal Q_i$
and using the properties of polylogarithms
(Eqs.~(A.4.2) and (A.2.7.6) of \citet{Lewin81}):
\begin{align}
\mathrm{Li}_3(-\mathrm e^x)-\mathrm{Li}_3(-\mathrm e^{-x})={}&{}-\frac{\pi^2x}6-\frac{x^3}{6},\\
\mathrm{Li}_4(-\mathrm e^x)+\mathrm{Li}_4(-\mathrm e^{-x})={}&{}-\frac{7\pi^4}{360}-\frac{\pi^2x^2}{12}-\frac{x^4}{24},\\
\mathrm{Li}_5(-\mathrm e^x)-\mathrm{Li}_5(-\mathrm e^{-x})={}&{}-\frac{7\pi^4x}{360}-\frac{\pi^2x^3}{36}-\frac{x^5}{120},\\
\mathrm{Li}_6(-\mathrm e^x)+\mathrm{Li}_6(-\mathrm e^{-x})={}&{}-\frac{31\pi^6}{15120}-\frac{7\pi^4x^2}{720}\notag\\
&{}-\frac{\pi^2x^4}{144}-\frac{x^6}{720},
\end{align}
after tedious but straightforward calculations, we obtain
\begin{align}
\label{eq:Rinonlin}
\mathcal R_i={}& \frac{8.86\times 10^{31}}{\unit{cm^{3}\,s}}
\sqrt[3]{\frac{Y_i\rho}{\rho_n}} \left(\frac{T}{\unit[10^9]{K}}\right)^5\notag\\
{}&\times\frac{17\pi^4x_i+10\pi^2x_i^3+x_i^5}{30},\\
\label{eq:Qinonlin}
\mathcal Q_i={}& \frac{1.22\times10^{25}}{\unit{erg^{-1}\,cm^{3}\,s}}
\sqrt[3]{\frac{Y_i\rho}{\rho_n}} \left(\frac{T}{\unit[10^9]{K}}\right)^6\notag\\
{}&\times\frac{\frac{457}{21}\pi^6 + 51\pi^4x_i^2 + 15\pi^2x_i^4 + x_i^6}{60}.
\end{align}

In order to simplify the code (i.e., in order to implement the implicit time evolution
with direct inversion rather than with an iterative procedure, see \citet{Camelio22code}), we linearize Eqs.~\eqref{eq:Rinonlin}
and \eqref{eq:Qinonlin} in $\Delta Y_i$ around the equilibrium configuration\footnote{Note that, for our EOS, (i) $\mathbb A^i$ is also
linear in $\Delta Y_i$, and (ii) $T=T^\req$ at first order in $\Delta Y_i$ (cf.~Eq.~(64) of \citet{Gavassino21bulk}).}:
\begin{align}
\label{eq:R}
\mathcal R_i\simeq{}& \frac{8.86\times 10^{31}}{\unit{cm^{3}\,s}}
\sqrt[3]{\frac{Y^\req_i\rho}{\rho_n}} \left(\frac{T^\req}{\unit[10^9]{K}}\right)^5
\frac{17\pi^4}{30}\frac{\mathbb A^i}{k_\mathrm{B}T^\req},\\
\label{eq:Q}
\mathcal Q_i\simeq{}& \frac{1.22\times10^{25}}{\unit{erg^{-1}\,cm^{3}\,s}}
\sqrt[3]{\frac{Y^\req_i\rho}{\rho_n}} \left(\frac{T^\req}{\unit[10^9]{K}}\right)^6\frac{457\pi^6}{1260}
\left(1 + \frac{\Delta Y_i}{3Y^\req_i}\right).
\end{align}

For modified beta reactions (modified Urca, see Sec.~2.2 of \citet{Haensel92}) the
calculations are similar but longer.
Following the procedure we detailed above\footnote{In addition,
$\mathrm{Li}_8(-\mathrm e^x)-\mathrm{Li}_8(-\mathrm e^{-x})=-{127\pi^8}/604800
+ \mathrm{O}(x^2)$.}, we compute the number ($\mathcal R_i^\mathrm{m}$) and energy
($\mathcal Q_i^\mathrm{m}$) modified Urca reaction rates for lepton species
$i$, linearized in $\Delta Y_i$ around the equilibrium configuration:
\begin{align}
\label{eq:R-murca}
\mathcal R_i^\mathrm{m}\simeq{}& \frac{5.91\times 10^{23}}{\unit{cm^{3}\,s}}
\sqrt[3]{\frac{Y^\req_i\rho}{\rho_n}}\left(\frac{T^\req}{\unit[10^9]{K}}\right)^7
\frac{367\pi^6}{63} \frac{\mathbb A^i}{k_\mathrm{B}T^\req},\\
\label{eq:Q-murca}
\mathcal Q_i^\mathrm{m}\simeq{}& \frac{8.15\times10^{16}}{\unit{erg^{-1}\,cm^{3}\,s}}
\sqrt[3]{\frac{Y^\req_i\rho}{\rho_n}} \left(\frac{T^\req}{\unit[10^9]{K}}\right)^8
\frac{11513\pi^8}{2520} \left(1 + \frac{\Delta Y_i}{3Y^\req_i}\right).
\end{align}
Only 3 particles plus one neutrino partake in a direct Urca reaction
(Eqs.~\eqref{eq:beta-minus-e}--\eqref{eq:beta-plus-e}, see also Eqs.~(1) and
(2) of \citet{Haensel92}), while 5 particles plus one neutrino partake in a
modified Urca reaction (Eqs.~(10) and (11) of \citet{Haensel92}).
As a consequence, in contrast to the direct Urca reactions
[cf.~Eqs.~\eqref{eq:R}--\eqref{eq:Q} with
Eqs.~\eqref{eq:R-murca}--\eqref{eq:Q-murca}], (i) the prefactor of the modified
Urca rates is smaller, (ii) the modified Urca rates depend on a larger power of
the temperature, and (iii) the modified Urca reactions are not kinematically
inhibited at lower density and temperature.

The aforementioned consequence (iii) is the main reason why modified Urca
reactions should be kept into account when studying neutron stars.
Moreover, theoretical estimates \cite{Alford18} and some numerical simulations
(e.g., \citet{Most22}) show that in some regimes (i.e., in neutron star
post-merger remnants) bulk viscosity from modified Urca reactions may provide
significant damping, although simulations with more sophisticated neutrino
transport and higher resolution do not find a significant imprint of
out-of-equilibrium effects on the gravitational wave emission \cite{Radice22,
Zappa23}.

We finally remark that the linearization of the reactions around equilibrium
[Eqs.~\eqref{eq:R}--\eqref{eq:Q-murca}] is not always correct in the cases
considered in the companion paper \cite{Camelio22b}.
Indeed, in our simulations $\mathbb A/k_\mathrm{B}T$ can become greater than
one, in particular for the electrons and in particular close to the surface.
However, also considering the other approximations employed (i.e., spherical
symmetry, a polytropic EOS, and only direct beta reactions), our aim is not to
obtain a quantitative description of a neutron star, but rather to study the
equivalence between different approaches to bulk viscosity in neutron stars.

\section{Hydrodynamic equations for bulk viscosity}
\label{sec:bulk-eqs}

In this section, we derive the hydrodynamic equations for the bulk stress in the
M\"uller-Israel-Stewart theories in radial gauge-polar slicing
coordinates in spherical symmetry (i.e.~Schwarzschild):
\begin{equation}
\label{eq:metric}
\mathrm dl^2= 
-\alpha^2(r,t)\mathrm dt^2 + X^2(r,t) \mathrm dr^2
+r^2\mathrm d\Omega^2,
\end{equation}
where $l$ is the proper time, $t$ and $r$ are respectively the time and radial
coordinates, $\mathrm d\Omega$ the angular element, $\alpha$ is the lapse and
$X$ is a metric function.

Using that the divergence of a vector
$A^\mu$ and of a scalar field $\varphi$ are (Appendix~A of \citet{OConnor10}):
\begin{align}
\sqrt{-g}\nabla_\mu A^\mu={}&\partial_\mu (\sqrt{-g}A^\mu),\\
\nabla_\mu \varphi={}&\partial_\mu\varphi,
\end{align}
and that in Schwarzschild coordinates the determinant of the metric $\sqrt{-g}$
and the fluid 4-velocity $u^\mu$ are:
\begin{align}
\sqrt{-g}={}&r^2\alpha X,\\
u^\mu={}&\left(\frac{W}\alpha,\frac{Wv}X,0,0\right),\\
\label{eq:w}
W={}&\left(1 - v^2\right)^{-1/2},
\end{align}
where $v=Xu^r/\alpha u^t$ is the physical velocity of the fluid
and $W=\alpha u^t$ a Lorentz-like factor, the bulk viscous
equations~\eqref{eq:bv-entropy}, \eqref{eq:hiscock-lindblom}, and
\eqref{eq:maxwell-cattaneo} become, respectively:
\begin{equation}
\label{eq:evol-bulk}
\partial_t (XW\rho s)
+\frac1{r^2}\partial_r\left(\alpha r^2 W\rho s v\right)=\alpha Xm_\rn\frac{\Pi^2}{\zeta T^\req},
\end{equation}
\begin{multline}
\label{eq:pi-evol-hl}
\partial_t(XW\Pi) + \frac1{r^2}\partial_r\left(r^2\alpha Wv\Pi\right) = -\frac{\alpha X\Pi}{\tau}\\
-\left(\frac1\chi-\frac{\Pi}2\right)\left(\partial_t (XW) +\frac1{r^2}\partial_r(r^2\alpha Wv)\right)\\
-\frac{\Pi W}2\left(X\partial_t\log\frac\chi{T^\req}+\alpha v\partial_r\log\frac\chi{T^\req}\right),
\end{multline}
\begin{multline}
\label{eq:pi-evol}
\partial_t(XW\Pi) + \frac1{r^2}\partial_r\left(r^2\alpha Wv\Pi\right) = -\frac{\alpha X\Pi}{\tau}\\
-\left(\frac1\chi-\Pi\right)\left(\partial_t (XW) +\frac1{r^2}\partial_r(r^2\alpha Wv)\right).
\end{multline}
Note that there are time
derivatives in the source of Eqs.~\eqref{eq:pi-evol-hl} and \eqref{eq:pi-evol},
which are a complication for the numerical implementation. However,
Eq.~\eqref{eq:evol-bulk} has bigger problems, since the inversion from the
entropy $s$ to the bulk stress $\Pi$ is degenerate [see Eq.~\eqref{eq:bv-s}].
In practice, we numerically implemented only Eqs.~\eqref{eq:pi-evol-hl} and
\eqref{eq:pi-evol} in the companion paper \cite{Camelio22b, Camelio22code}, since we did not
manage to evolve Eq.~\eqref{eq:evol-bulk} in a stable manner.

For our choice of EOS and reaction rates, Eqs~\eqref{eq:def-xi}, \eqref{eq:def-zeta}, \eqref{eq:tau-multispecies},
and \eqref{eq:a2pi} become:
\begin{align}
\Xi_{ab}={}& \mathrm{diag}(\Xi_\re,\Xi_\mu),\\
\Xi_i={}& \frac{8.86\times10^{31}}{\unit{cm^{3}\,s}}\sqrt[3]{\frac{Y^0_i\rho^2}{\rho^2_n}}
\left(\frac{T}{\unit[10^9]{K}}\right)^5
\frac{17\pi^4}{30k_BT},\\
\zeta={}&n^2\frac{\rho^2}{\rho_n^2}\sum_i\frac{(Y^0_i)^2}{\Xi_i},\\
\tau={}&\frac{n}{2m_\rn}\times\sum_i{\frac{(Y_i^0)^2}{\Xi_i}}\div\sum_i{k_i(Y_i^0)^2},\\
\Pi={}&-2\rho\sum_i k_i Y^\req_i \Delta Y_i,
\end{align}
where $i=\{\re,\mu\}$.
Note that, with our model, the second term of Eq.~\eqref{eq:effective-cs2} does not
diverge for $\rho\to0$.

\section{Conclusions}
\label{sec:conclusions}

With this paper we aim to clarify how different approaches to bulk viscosity
can be implemented in practice.
In order to do so, we cast the equations of the multi-component fluid and of
two M\"uller-Israel-Stewart theories (Hiscock-Lindblom and Maxwell-Cattaneo) in
conservative form in radial gauge, polar slicing coordinates and spherical
symmetry.

The novel results presented in this paper are an example of
inversion from the M\"uller-Israel-Stewart formulation to the multi-component
fluid one (Sec.~\ref{ssec:is2carter}), the extension of the
M\"uller-Israel-Stewart theories to include the energy loss due to the
reactions (Sec.~\ref{ssec:q}), the specialization of the
M\"uller-Israel-Stewart equations to the radial gauge-polar slicing coordinates
in spherical symmetry (Sec.~\ref{sec:bulk-eqs}), and the computation of the
signal propagation speed for the Hiscock-Lindblom theory in the non-linear
regime (Appendix~\ref{sec:signal-speed}).  We also propose a simple but
physically motivated equation of state (Sec.~\ref{ssec:eos}) and an analytical
form for the reaction rates (Sec.~\ref{ssec:rates}).

This paper establishes the theoretical foundations for a companion paper
\cite{Camelio22b}, in which we implement a one-dimensional general relativistic
hydrodynamic code \cite{Camelio22code} to study bulk viscosity in neutron stars.
As numerically confirmed in the companion paper \citep{Camelio22b}, the
M\"uller-Israel-Stewart and the multi-component fluid formulations are
equivalent for small thermodynamic perturbations \citep{Gavassino21bulk}.  In
our opinion, when bulk viscosity is due to particle reactions, it is more easy,
convenient, and accurate to directly evolve the multi-component fluid rather
than implementing bulk viscosity with a perturbative formulation such as the
M\"uller-Israel-Stewart theories, which is often adopted for binary neutron
star mergers.
Note that the multi-component fluid is already the normal approach in
core-collapse supernova simulations (e.g., \citet{OConnor10}), where knowing the
exact composition of the matter is considered very important to correctly
determine the evolution of the supernova and its observables, like the neutrino
luminosity \cite{Burrows88}.

Moreover, even if bulk viscosity is not due to particle reactions, it is
possible to reformulate the problem in terms of a multi-component fluid, whose
additional fictitious particle species (also called reaction coordinate)
effectively accounts for bulk viscosity (see Sec.~\ref{ssec:is2carter}).

This paper and its companion \cite{Camelio22b} make many strong approximations:
spherical symmetry, a polytropic EOS, the focus on direct Urca reactions over
the modified ones, and the linearization of the reactions in $\Delta
\mu/k_\mathrm{B}T$.
These approximations have been made not to clutter the numerical implementation
with technical details that would obscure our main aim: the comparison between
the multi-component and the bulk stress descriptions of viscosity in the
context of neutron stars.
As a consequence, our model should be considered as a playground to experiment
with bulk viscosity and obtain qualitative results; more physics should be
added in order to obtain quantitative physical predictions.

As a final remark, we point out that the correspondence between the
multi-component fluid and bulk stress can be used also to describe
superfluidity \cite{superbulk_PRD_2022} and in contexts different from neutron
stars, as in cosmology \cite{Hiscock91, Zakari93, Maartens95} and particle
physics \cite{DelZanna13}.

\begin{acknowledgments}
This work was supported by the Polish National Science Centre (NCN) grant number OPUS~2019/33/B/ST9/00942.
SB acknowledges support by the EU H2020 under ERC Starting Grant, no.~BinGraSp-714626.
LG is partially supported by a Vanderbilt's Seeding Success Grant.
\end{acknowledgments}

\bibliographystyle{unsrtnat}
\bibliography{paper20230519.bbl}

\appendix

\section{Comparing the speeds of sound}
\label{sec:cs-vs-cs}

\subsection{Some useful thermodynamic relations}
\label{ssec:thermo-useful}

Here, we derive Eqs.~\eqref{eq:useful1} and \eqref{useful2}, which will
be used in the next subsections.

At equilibrium, the internal specific energy must be in a minimum \cite{Callen_book,GavassinoGibbs2021}, for
constant $\rho$ and $s$. Hence, the symmetric matrix
\begin{equation}
u^{ab}:= \dfrac{\partial^2 u}{\partial Y_a \partial Y_b}
\bigg|_{\rho,s,\{Y_i\}_{i\neq a,b}} = -\dfrac{1}{m_\rn} \dfrac{\partial
\mathbb{A}^a}{\partial Y_b}\bigg|_{\rho,s,\{Y_i\}_{i\neq b}}
\end{equation}
is non-negative definite close to equilibrium.
If we differentiate the equilibrium condition
\begin{equation}
\mathbb{A}^a (\rho,s, \{Y_i^\req(\rho,s)\}_i )=0\qquad\forall a,
\end{equation}
with respect to $\rho$ we obtain (repeated species indices are summed with the Einstein convention):
\begin{equation}
\dfrac{\partial \mathbb{A}^a}{\partial \rho}\bigg|_{s,\{Y_i\}_i} = m_\rn
u^{ab} \dfrac{\partial Y_b^\req}{\partial \rho} \bigg|_s ,
\end{equation}
which implies, since $u^{ab}$ is non-negative definite,
\begin{equation}
\label{eq:useful1}
\dfrac{\partial Y_a^\req}{\partial \rho} \bigg|_s\dfrac{\partial
\mathbb{A}^a}{\partial \rho}\bigg|_{s,\{Y_i\}_i} = m_\rn  u^{ab}
\dfrac{\partial Y_a^\req}{\partial \rho} \bigg|_s \dfrac{\partial
Y_b^\req}{\partial \rho} \bigg|_s \geq 0.
\end{equation}

The differentiation of the pressure with respect to $Y_a$ at constant
$\epsilon$ is connected to that at constant $\rho$ by this formula:
\begin{multline}
\label{useful3}
\dfrac{\partial p}{\partial Y_a} \bigg|_{\epsilon,s,\{Y_i \}_{i\neq a}}=
\dfrac{\partial p}{\partial Y_a} \bigg|_{\rho,s,\{Y_i \}_{i\neq a}}\\
+\dfrac{\partial p}{\partial \rho}\bigg|_{s,\{Y_i\}_i}  \dfrac{\partial
\rho}{\partial Y_a} \bigg|_{\epsilon,s,\{Y_i \}_{i\neq a}}.
\end{multline}
From the thermodynamic differential
\begin{equation}
\mathrm d\epsilon = \dfrac{\rho T}{m_\rn}\mathrm  ds +\dfrac{\epsilon +p}{\rho} \mathrm
d\rho -  \dfrac{\rho \mathbb{A}^a}{m_\rn}\mathrm dY_a
\end{equation}
we obtain
\begin{equation}
\dfrac{\partial \rho}{\partial Y_a} \bigg|_{\epsilon,s,\{Y_i \}_{i\neq a}} =
\dfrac{\rho^2 \mathbb{A}^a}{m_\rn (\epsilon+p)},
\end{equation}
and from Eq.~\eqref{eq:eos-th} we obtain the Maxwell relation
\begin{equation}
\dfrac{\partial p}{\partial Y_a} \bigg|_{\rho,s,\{Y_i \}_{i\neq a}} =
-\dfrac{\rho^2}{m_\rn} \dfrac{\partial \mathbb{A}^a}{\partial
\rho}\bigg|_{s,\{Y_i\}_i},
\end{equation}
such that Eq.~\eqref{useful3} becomes
\begin{equation}
\label{useful2}
\dfrac{\partial p}{\partial Y_a} \bigg|_{\epsilon,s,\{Y_i \}_{i\neq a}} =
\frac{\rho^2}{m_\rn}\left(\dfrac{\mathbb{A}^a}{\epsilon+p}\dfrac{\partial p}{\partial
\rho}\bigg|_{s,\{Y_i\}_i}
 -\dfrac{\partial \mathbb{A}^a}{\partial \rho}\bigg|_{s,\{Y_i\}_i}\right).
\end{equation}

\subsection{Ultraviolet vs infrared}
\label{sec:uv-vs-ir}

The differentiation of the pressure with respect to $\epsilon$ at constant $\{\mathbb A^i\}_i$
is connected to that at constant $\{Y_i\}_i$ by this formula:
\begin{equation}
\dfrac{\partial p}{\partial \epsilon}\bigg|_{s,\{\mathbb{A}^i\}_i} = \dfrac{\partial
p}{\partial \epsilon}\bigg|_{s,\{Y_i\}_i} + \dfrac{\partial p}{\partial
Y_a}\bigg|_{\epsilon,s,\{ Y_i \}_{i\neq a}} \dfrac{\partial Y_a}{\partial
\epsilon}\bigg|_{s,\{\mathbb{A}^i\}_i},
\end{equation}
which, using Eqs.~\eqref{eq:cs2-uv} and \eqref{useful2}, becomes 
\begin{multline}
c_\mathrm{s,uv}^2 = \dfrac{\partial p}{\partial
\epsilon}\bigg|_{s,\{\mathbb{A}^i\}_i}+ 
\dfrac{\rho^2}{m_\rn}\dfrac{\partial \rho}{\partial
\epsilon}\bigg|_{s,\{\mathbb{A}^i\}_i} \dfrac{\partial Y_a}{\partial
\rho}\bigg|_{s,\{\mathbb{A}^i\}_i}\\
\times \left[ \dfrac{\partial \mathbb{A}^a}{\partial
\rho}\bigg|_{s,\{Y_i\}_i} \! \! \! \! - \dfrac{ \mathbb{A}^a}{
(\epsilon+p)}\dfrac{\partial p}{\partial \rho}\bigg|_{s,\{Y_i\}_i}  \right].\\
\end{multline}
If we evaluate this formula at equilibrium, it reduces to
\begin{align}
\label{useful4}
(c^\req_\mathrm{s,uv})^2 ={}& c_\mathrm{s,ir}^2 + \dfrac{\rho^3}{m_\rn
(\epsilon+p^\req)} \dfrac{\partial Y_a^\req}{\partial \rho}\bigg|_{s}
\dfrac{\partial \mathbb{A}^a}{\partial \rho}\bigg|_{s,\{Y_i\}_i}^\req\\
\geq{}& c_\mathrm{s,ir}^2,\notag
\end{align}
where the inequality follows from Eq.~\eqref{eq:useful1}.

\subsection{M\"uller-Israel-Stewart vs multi-constituent fluid}
\label{ssec:is-vs-mcf}

At equilibrium, Eqs.~\eqref{eq:cs2-ir} and \eqref{eq:csir-bulk} are clearly the
same, because $s^\req =s$ and $p^\req =p$. Hence, comparing
Eqs.~\eqref{eq:effective-cs2} and \eqref{useful4}, and recalling
Eq.~\eqref{eq:useful1}, we obtain that the equilibrium ultraviolet speed of
sound of the multi-component fluid coincides with that of the bulk
stress theory if
\begin{equation}
\chi^{-1} = \rho^3  u^{ab} \dfrac{\partial Y_a^\req}{\partial \rho} \bigg|_s \dfrac{\partial Y_b^\req}{\partial \rho} \bigg|_s .
\end{equation}
Combining this equation with Eqs.~\eqref{eq:bv-tau} and \eqref{eq:def-zeta}, we
finally obtain Eq.~\eqref{eq:tau-multispecies}.

\section{Propagation speed of a signal in the Hiscock-Lindblom theory}
\label{sec:signal-speed}

In this section we compute the characteristic speeds of the
Hiscock-Lindblom theory in the non-linear regime, following \citet{Bemfica19}.
The hydrodynamic equations can be written as \citep{Bemfica19}:
\begin{align}
\label{eq:bem1}
u^\mu\nabla_\mu\epsilon + \hden\nabla_\mu u^\mu={}&0,\\
\label{eq:bem2}
a_1\Delta^{\mu\nu}\nabla_\mu\epsilon + \hden u^\mu\nabla_\mu u^\nu
+a_2\Delta^{\mu\nu}\nabla_\mu n+\Delta^{\mu\nu}\nabla_\mu\Pi={}&0,\\
\label{eq:bem3}
n \nabla_\mu u^\mu + u^\mu\nabla_\mu n={}&0,\\
\label{eq:hl}
b_1u^\mu\nabla_\mu\epsilon + \tilde\zeta\nabla_\mu u^\mu
+b_2u^\mu\nabla_\mu n + \tau u^\mu\nabla_\mu\Pi + \Pi={}&0,
\end{align}
where in the last equation we have generalized Eq.~(4) of \citet{Bemfica19}
to the Hiscock-Lindblom theory and:
\begin{align}
\hden={}&\epsilon+p_\req+\Pi,\\
\Delta^{\mu\nu}={}&g^{\mu\nu} + u^\mu u^\nu,\\
a_1={}&\left.\frac{\partial p_\req}{\partial\epsilon}\right|_n,\\
a_2={}&\left.\frac{\partial p_\req}{\partial n}\right|_\epsilon,\\
\tilde\zeta={}&\zeta + \frac{\tau\Pi}2,\\
b_1={}&\frac{\zeta\Pi T_\req}2\left.\frac{\partial \chi/ T_\req}{\partial\epsilon}\right|_n,\\
b_2={}&\frac{\zeta\Pi T_\req}2\left.\frac{\partial \chi/ T_\req}{\partial n}\right|_\epsilon.
\end{align}
In order to compute the characteristics, we need first to write Eqs.~\eqref{eq:bem1}--\eqref{eq:hl}
in the form
\begin{equation}
\mathcal M^\mu\nabla_\mu\psi + \mathcal N\psi=0,
\end{equation}
where $\mathcal M^\mu$ and $\mathcal N$ are matrices and
\begin{equation}
\psi=(\epsilon,u^\mu,n,\Pi).
\end{equation}
The matrix $\mathcal M^\mu$ is
\begin{equation}
\label{eq:mmatrix}
\mathcal M^\mu =
\begin{bmatrix}
u^\mu & \hden\delta_\nu^\mu & 0 & 0 \\
a_1\Delta^{\nu\mu} & \mathbb I_{4\times4}\hden u^\mu & a_2\Delta^{\nu\mu} & \Delta^{\nu\mu} \\
0 & n\delta_\nu^\mu & u^\mu & 0\\
b_1u^\mu & \tilde\zeta\delta_\nu^\mu & b_2u^\mu & \tau u^\mu
\end{bmatrix}.
\end{equation}
The characteristics of the system are determined\footnote{In order to obtain
Eq.~\eqref{eq:chars} one can directly compute the determinant of $\mathcal
M^\mu\xi_\mu$, noting that the second column/row of Eq.~\eqref{eq:mmatrix} are
actually 4 columns/rows, or one can subtract column 4 multiplied by $b_1/\tau$
(resp.~by $b_2/\tau$) to column 1 (resp.~column 3) and use the result of
\citet{Bemfica19} with $\zeta\to\tilde\zeta$, $\alpha_1\to a_1 - b_1/\tau$, and
$\alpha_2\to a_2 - b_2/\tau$.} by the $\xi_\mu$ such that $\det(\mathcal
M^\mu\xi_\mu)=0$. We have that:
\begin{multline}
\label{eq:chars}
\det(M^\mu\xi_\mu)=(u^\mu\xi_\mu)^5\hden^4\tau\Bigg[(u^\mu\xi_\mu)^2\\
- \left( a_1 - \frac{b_1}{\tau} + \frac{\tilde\zeta}{\tau\hden} +
  \frac{n}{\hden}\left(a_2-\frac{b_2}{\tau}\right)
  \right)\Delta^{\mu\nu}\xi_\mu\xi_\nu\Bigg],
\end{multline}
and therefore the propagation speed of a signal (i.e., the effective sound speed) squared is:
\begin{equation}
\label{eq:hl-propagation-speed}
c_{s,\mathrm{uv}}^2=a_1 - \frac{b_1}{\tau} + \frac{\tilde\zeta}{\tau\hden} +
\frac{n}{\hden}\left(a_2-\frac{b_2}{\tau}\right).
\end{equation}
Close to equilibrium we obtain, as for Maxwell-Cattaneo, Eq.~\eqref{eq:effective-cs2}.

\end{document}